\documentclass[twoside,twocolumn,9pt]{article}
\usepackage{extsizes}
\usepackage[super,sort&compress,comma]{natbib} 
\usepackage[version=3]{mhchem}
\usepackage[left=1.5cm, right=1.5cm, top=1.785cm, bottom=2.0cm]{geometry}
\usepackage{balance}
\usepackage{mathptmx}
\usepackage{sectsty}
\usepackage{graphicx} 
\usepackage{lastpage}
\usepackage[format=plain,justification=justified,singlelinecheck=false,font={stretch=1.125,small,sf},labelfont=bf,labelsep=space]{caption}
\usepackage{float}
\usepackage{fancyhdr}
\usepackage{fnpos}
\usepackage[english]{babel}
\addto{\captionsenglish}{%
  
}
\usepackage{array}
\usepackage{droidsans}
\usepackage{charter}
\usepackage[T1]{fontenc}
\usepackage[usenames,dvipsnames]{xcolor}
\usepackage{setspace}
\usepackage[compact]{titlesec}
\usepackage{hyperref}

\usepackage{amsmath,amssymb}

\usepackage{epstopdf}

\definecolor{cream}{RGB}{222,217,201}

\begin{document}

\pagestyle{fancy}
\thispagestyle{plain}
\fancypagestyle{plain}{
\renewcommand{\headrulewidth}{0pt}
}

\makeFNbottom
\makeatletter
\renewcommand\LARGE{\@setfontsize\LARGE{15pt}{17}}
\renewcommand\Large{\@setfontsize\Large{12pt}{14}}
\renewcommand\large{\@setfontsize\large{10pt}{12}}
\renewcommand\footnotesize{\@setfontsize\footnotesize{7pt}{10}}
\makeatother

\renewcommand{\thefootnote}{\fnsymbol{footnote}}
\renewcommand\footnoterule{\vspace*{1pt}%
\color{cream}\hrule width 3.5in height 0.4pt \color{black}\vspace*{5pt}} 
\setcounter{secnumdepth}{5}

\makeatletter 
\renewcommand\@biblabel[1]{#1}            
\renewcommand\@makefntext[1]%
{\noindent\makebox[0pt][r]{\@thefnmark\,}#1}
\makeatother 
\renewcommand{\figurename}{\small{Fig.}~}
\sectionfont{\sffamily\Large}
\subsectionfont{\normalsize}
\subsubsectionfont{\bf}
\setstretch{1.125} 
\setlength{\skip\footins}{0.8cm}
\setlength{\footnotesep}{0.25cm}
\setlength{\jot}{10pt}
\titlespacing*{\section}{0pt}{4pt}{4pt}
\titlespacing*{\subsection}{0pt}{15pt}{1pt}

\fancyfoot{}
\fancyfoot[LO,RE]{\vspace{-7.1pt}\includegraphics[height=9pt]{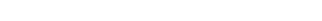}}
\fancyfoot[CO]{\vspace{-7.1pt}\hspace{13.2cm}\includegraphics{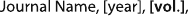}}
\fancyfoot[CE]{\vspace{-7.2pt}\hspace{-14.2cm}\includegraphics{head_foot/RF}}
\fancyfoot[RO]{\footnotesize{\sffamily{1--\pageref{LastPage} ~\textbar  \hspace{2pt}\thepage}}}
\fancyfoot[LE]{\footnotesize{\sffamily{\thepage~\textbar\hspace{3.45cm} 1--\pageref{LastPage}}}}
\fancyhead{}
\renewcommand{\headrulewidth}{0pt} 
\renewcommand{\footrulewidth}{0pt}
\setlength{\arrayrulewidth}{1pt}
\setlength{\columnsep}{6.5mm}
\setlength\bibsep{1pt}

\makeatletter 
\newlength{\figrulesep} 
\setlength{\figrulesep}{0.5\textfloatsep} 

\newcommand{\topfigrule}{\vspace*{-1pt}%
\noindent{\color{cream}\rule[-\figrulesep]{\columnwidth}{1.5pt}} }

\newcommand{\botfigrule}{\vspace*{-2pt}%
\noindent{\color{cream}\rule[\figrulesep]{\columnwidth}{1.5pt}} }

\newcommand{\dblfigrule}{\vspace*{-1pt}%
\noindent{\color{cream}\rule[-\figrulesep]{\textwidth}{1.5pt}} }

\makeatother

\twocolumn[
  \begin{@twocolumnfalse}
{\includegraphics[height=30pt]{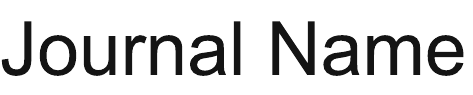}\hfill\raisebox{0pt}[0pt][0pt]{\includegraphics[height=55pt]{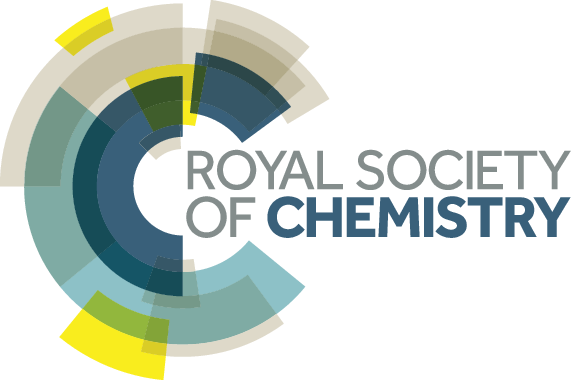}}\\[1ex]
\includegraphics[width=18.5cm]{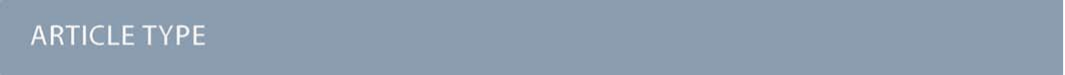}}\par
\vspace{1em}
\sffamily
\begin{tabular}{m{4.5cm} p{13.5cm} }

\includegraphics{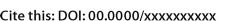} & \noindent\LARGE{\textbf{Allotropic Ga$_2$Se$_3$/GaSe nanostructures grown by van der Waals epitaxy: Narrow exciton lines and single-photon emission$^\dag$}} \\
\vspace{0.3cm} & \vspace{0.3cm} \\

 & \noindent\large{Maxim Rakhlin, Sergey Sorokin, Aidar Galimov, Ilya Eliseyev, Valery Davydov, Demid~\mbox{Kirilenko}, Alexey Toropov, and Tatiana Shubina$^{\ast}$} \\

\includegraphics{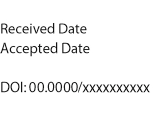} & \noindent\normalsize{The ability to emit narrow exciton lines, preferably with a clearly defined polarization, is one of the key conditions for the use of nanostructures based on III-VI monochalcogenides and other layered crystals in quantum technology to create non-classical light. Currently, the main method of their formation is exfoliation followed by strain and defect engineering. A factor limiting the use of epitaxy is the presence of different phases in the grown films. In this work, we show that control over their formation makes it possible to create structures with the desired properties. We propose Ga$_2$Se$_3$/GaSe nanostructures by van der Waals epitaxy with a high VI/III flux ratio as a source of narrow exciton lines. Actually, these nanostructures are a combination of allotropes: GaSe and Ga$_2$Se$_3$, consisting of the same atoms in different arrangements. The energy position of the narrow lines is determined by the quantum confinement in Ga$_2$Se$_3$ inclusions of different sizes in the GaSe matrix, similar to quantum dots, and their linear polarization is due to the ordering of Ga vacancies in a certain crystalline direction in Ga$_2$Se$_3$.  Such nanostructures exhibit single-photon emission with second-order correlation function $g^{(2)}(0)\sim$0.1 at 10 K that makes them promising for quantum technologies. 
}
\\

\end{tabular}

 \end{@twocolumnfalse} \vspace{0.6cm}

     ]
    
    \renewcommand*\rmdefault{bch}\normalfont\upshape
    \rmfamily
    \section*{}
    \vspace{-1cm}


\footnotetext{\textit{$^{a}$~Ioffe  Institute, St.~Petersburg, 194021, Russia. E-mail:shubina@beam.ioffe.ru}}

\footnotetext{\dag~Electronic Supplementary Information (ESI) available: 
See DOI: 00.0000/00000000.}




\section{Introduction}
The last decades can be characterized as an era of two-dimensional (2D) physics, when layered anisotropic semiconductors -- transition  metal dichalcogenides  and group III metal monochalcogenides, studied since the middle of the last century, are being discovered from a new perspective. Monochalcogenides such as GaSe, GaS, and InSe are attracting much attention as parent materials for 2D and few-layer structures promising for applications in nanoelectronics (field-effect transistors), photovoltaics, sensors, and nonlinear optics \cite{Bourdon, Guo, Sato, Cai, Wang2020}. Strain and defect engineering has been used to realize states similar to quantum-dots in 2D structures that emit narrow exciton lines and can act as single photon sources \cite{Vasconcellos2022}. In particular, the ability to generate single-photon emission from defects in thin gallium selenide flakes exfoliated from a single crystal and transferred to a waveguide was demonstrated by Tonndorf et al.\cite{TonndorfNano}. 

Generally, GaSe, which is one of the most studied monochalcogenide III–VI semiconductors \cite{belenkii, Sarkar2020}. Unlike transition metal dichalcogenides, bulk GaSe has practically degenerate indirect and direct band gaps, which greatly simplifies the task of forming efficiently emitting structures \cite{Sun2018, Lai2022}. Published low-temperature photoluminescence (PL) spectra of bulk GaSe are inhomogeneous and contain a number of lines that were attributed to free excitons and excitons bound at various defects \cite{Capozzi1989} and various polytypes that differ from each other in the sequence of stacking in the unit cell \cite{Ueno1997, Liu2015}. When the number of monolayers is reduced to approximately six, GaSe emission disappears due to the very weak absorption of exciting light in indirect band gap structures \cite{DelPozo2015}.

A GaSe monolayer consists of four atomic layers, that is, a tetralayer, where two layers of Ga atoms are located between two layers of chalcogen atoms (Se-Ga-Ga-Se). This architecture allows the sliding of the Se sublayer \cite{Li2023} and, with an imbalance of the atomic composition, the formation of another structural phase - Ga$_2$Se$_3$. In fact, Ga$_2$Se$_3$ is an allotrope of GaSe, consisting of the same Ga and Se atoms, but in a different arrangement. This compound is called defective zinc blende, which, when Ga vacancies are ordered, has either a monoclinic (mono) or orthorhombic (ortho) crystal structure (Fig. S3 in the Supplementary). 

The electronic properties of ordered Ga$_2$Se$_3$ are almost independent on the particular crystal structure (mono or ortho) \cite{Peressi1998}. Unlike GaSe, in Ga$_2$Se$_3$ one third of the gallium positions are not filled and represent structural vacancies that are not located randomly, but along every third line in a certain direction, forming a “superstructure”, that is, an expanded unit cell \cite{Nakayama1997} (Fig. S4 in Supplementary). As a result, the PL turns out to be strongly linearly polarized \cite{Okamoto1994}. Similar phenomenon of optical anisotropy was observed in Ga$_2$S$_3$ as well \cite{Ho2014}. 

The measured band gap values of mono and ortho Ga$_2$Se$_3$ are characterized by a large scatter and depend both on the methods of film preparation and on the spectroscopic techniques \cite{Park1989, Okamoto1994, Morley1996,Emde1996, Lovejoy2010}. However, the difference in band gap between these phases is always within 0.5 eV and the mono phase has a higher energy, which is in good agreement with theoretical calculations \cite{Nakayama1997, Huang2013}. 

\begin{figure*}[t]
\includegraphics[width=\linewidth]{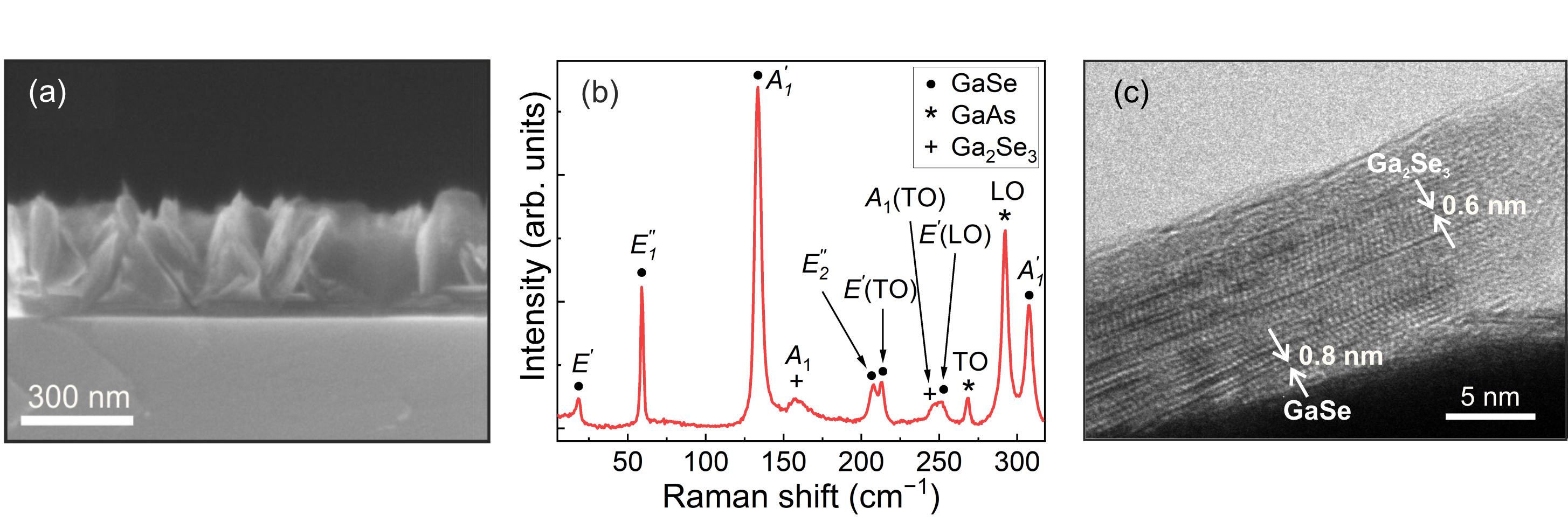}
\caption{SEM image of the cross section (a) and the unpolarized Raman spectrum (b) of a Ga$_2$Se$_3$/GaSe nanostructure grown by van der Waals MBE on GaAs (001). (c) High-resolution TEM images of Ga$_2$Se$_3$/GaSe nanostructures in cross-sectional geometry. 
\label{figSR}}
\end{figure*}

Despite the unique properties of bulk Ga$_2$Se$_3$, the Ga$_2$Se$_3$/GaSe nanostructures, which may be similar to quantum dots (QDs), have not yet been created and studied. The exfoliation technology typically used to fabricate 2D nanostructures is not suitable for creating such samples. Potentially, they can be formed by the epitaxial method. Molecular beam epitaxy (MBE) has certain advantages over other technologies due to the use of high-purity materials, ultra-high vacuum in the chamber and the ability to realize layer thickness at the atomic level \cite{Toropov2020}. The MBE in the van der Waals growth mode was proposed back in the 1990s \cite{Ohuchi1990, Koma1999} as a method for successful epitaxial growth even with strong layer-substrate lattice mismatch and when the crystal structures of constituents  are different. However, despite some promising results \cite{Yuan2015, Sorokin2019, Sorokin2020, Ohtake2021}, van der Waals epitaxy of both mono- and dichalcogenides is still under development \cite{Singh2022}.

In this paper, we report the successful formation of Ga$_2$Se$_3$/GaSe nanostructures using van der Waals MBE, in which an increase in the Se/Ga ratio with increasing temperature promoted the creation of Ga vacancies. Structural characterization showed that the grown samples contain ultra-small Ga$_2$Se$_3$ inclusions. Using micro-PL spectroscopy ($\mu$-PL), highly linearly polarized narrow emission lines of excitons (width $\sim$1 meV) were observed. They turned out to be strongly linearly polarized, as would be expected from the ordering of Ga vacancies. A $\sim$60$^\circ$ shift is observed between sets of polarized lines belonging to mono and ortho inclusions. The short decay times ($\sim$400 ps) of the lines are consistent with the direct band gap of Ga$_2$Se$_3$, and their quenching with increasing temperature is consistent with the use of GaSe as a barrier. The recorded single-photon emission of narrow lines with g$^{(2)}(0)\sim$0.1 indicates the presence of clear quantum levels in QD-like Ga$_2$Se$_3$/GaSe nanostructures.

\section{Sample growth and characterization}
\subsection{Van der Waals epitaxy}
Controlling the stoichiometry of van der Walls III-VI monochalcogenide films, tending to spontaneous formation of a large number of polytypes, requires a developed approach that takes into account relatively low growth temperatures as well as the possibility of the formation of transitional submonolayers at the heterointerface between the substrate and the layer \cite{Komkov2020, Sorokin2020, Dai1998}. 

The sample under study was grown on a GaAs(001) substrate using a two-chamber MBE setup. First, a 200 nm thick GaAs buffer layer was grown in a separate III-V chamber, then transferred to the III-VI chamber through vacuum, and the growth of GaSe film was started at a substrate temperature of $T_{S}$=400$^\circ$C. Such a relatively low value of $T_{S}$ provides the van der Waals growth mode with the orientation of the c axis of the growing GaSe layer perpendicular to the substrate surface, which ensures the formation of a sharp GaSe/GaAs(001) interface \cite{Sorokin2020}. After the growth of the first $\sim$ 4 nm, the epitaxy temperature was raised to $T_{S}$$\sim$500$^\circ$C and then remained unchanged until the end of the growth run. One of the main reasons for introducing this step is that GaSe layers grown at high $T_{S}$ demonstrate efficient near band-edge PL in contrast to the films grown at low $T_{S}$ \cite{Sorokin2020}. On the other hand, higher $T_{S}$ requires an increase in the VI/III ratio to avoid the appearance of Ga droplets on the growth surface. In this regard, it should be noted that the conditions established for the MBE growth of single-crystal Ga$_2$Se$_3$ layers do not differ much from that for GaSe: $T_{S}$ $\sim$ 450-550$^\circ$C (i.e., nearly the same), VI/III ratio is 15 and higher (i.e, only a few times higher in contrast to GaSe) if both standard Ga and Se sources as well as GaP(001) or GaAs(001) substrates are used \cite{Teraguchi1991, Teraguchi2, Okamoto1995, Okamoto1993}. In case of using valve cracking cell as Se source, the VI/III ratio required to maintain the Se-rich conditions during the growth of GaSe film is significantly higher due to the lower selenium sticking coefficient, and also depends on substrate temperature. Such a high VI/III flux ratio, as well as relatively low growth temperatures ($T_{S}$$\sim$500$^\circ$C) can apparently promote the formation of a vacancy-ordered Ga$_2$Se$_3$ phase inclusions in the growing film \cite{Huang2013, Okamoto1995}. Proof of this assumption is the existance of a peak associated with Ga$_2$Se$_3$ phase in the X-ray diffraction curves of GaSe layers grown at high VI/III flux ratio ($\sim$ 30-40) at $T_{S}$$\sim$400$^\circ$C using a Se valve cracking cell (see Supplementary). Thus, one can expect the appearance of a small amount of Ga$_2$Se$_3$ phase inclusions in GaSe layers grown in two-stage mode using the above parameters.

Scanning electron microscopy (SEM) cross-sectional images show that the total thickness of the grown film reached $\sim$ 0.36 $\mu$m (Fig. \ref{figSR})a. The film surface morphology is rough, enriched with trigonal or tetrahedral ridges, which results from the increase in temperature at the second stage. Further details of van der Waals epitaxy of GaSe can be found in the Methods section.

\subsection{Raman studies}

GaSe crystals consisting of Ga-Se-Se-Ga tetralayers exist as $\beta$, $\gamma$, $\delta$, and $\epsilon$ \cite{Nagel, Shluter, Segura2018} polytypes, which differ in mutual arrangement of tetralayers in a unit cell. To determine the main polytype of the sample grown by van der Waals epitaxy, we carried out Raman studies in a wide frequency range. The $\epsilon$ polytype occurs most frequently in epitaxial thin films. The high-frequency ($\omega>50$ cm$^{-1}$) Raman spectra of bulk $\epsilon$-GaSe ($D_{3h}$ point symmetry group) contain lines at 60, 134, 214, 252, and 308 cm$^{-1}$ named E$^{''}$, A$^{1}_{'}$, E$^{'}$( TO), E$^{'} $(LO), and A$^{'}_{1}$ respectively \cite{Hoff1975, Lei2013}. As seen in Figure \ref{figSR}b, the Raman spectrum obtained from the grown sample contains all these high-frequency lines. In addition, modes in the low-frequency region ($\omega<50$ cm$^{-1}$) make it possible to accurately determine the polytype \cite{Lei2013}. Here we observe one peak at 19 cm$^{-1}$, which convincingly confirms the predominance of the $\epsilon$ polytype of GaSe in the studied thin film.

The vacancy-ordered Ga$_2$Se$_3$ can exist in both monoclinic and orthorhombic modifications, whose electronic properties are similar \cite{Peressi1998}. We denote these modifications as “mono” and “ortho”, although some studies used the designations $\beta$-Ga$_2$Se$_3$ for mono and $\gamma$-Ga$_2$Se$_3$ for ortho \cite{Ho2020}. In the Raman spectrum of the sample under study, the 155 cm$^{-1}$ and $\sim$250 cm$^{-1}$ modes are observed, which were previously registered in Ga$_2$Se$ _3$ \cite{Finkman1975, Kolodziejczyk1992, Yamada1992}. The peak at 155 cm$^{-1}$ can be attributed to the A$_1$ mode of the vacancy-ordered phase \cite{Finkman1975}. The broadened shape of the peak indicates a change in the crystallinity of the Ga$_2$Se$_3$ inclusions; a similar shape was observed in \cite{Yamada1992} for specimens grown at relatively low VI/III flux ratios. In the region of 250 cm$^{-1}$, a doublet is observed, consisting of two peaks: a weak E$^{'}$(LO) peak of bulk GaSe (252 cm$^{-1}$) and a peak at 245 cm$^{-1}$. It was suggested that the latter may be due to short-range interactions in the Ga$_2$Se$_3$ unit cell \cite{Emde1995}. Thus, the observed combination of modes in the Raman spectrum point to coexistence of $\epsilon$-GaSe and vacancy-ordered Ga$_2$Se$_3$. 

In addition, several lines in the Raman spectrum in Figure \ref{figSR}b arise from growth on the GaAs substrate and ridge morphology. For example, two additional lines at 268 and 291 cm$^{-1}$ correspond to transverse (TO) and longitudinal (LO) optical phonons of the GaAs substrate. A distinct doublet at 200-220 cm$^{-1}$ is a sign of rough morphology \cite{Sorokin2020}, when the appearance of the low-frequency E$^{''}_{2}$ mode in GaSe became possible due to the inclination of the ridge face depending on the incident light \cite{Irwin1973, Hoff1975}. 


\begin{figure*}[t]
\centering
\includegraphics[width=0.9\linewidth]{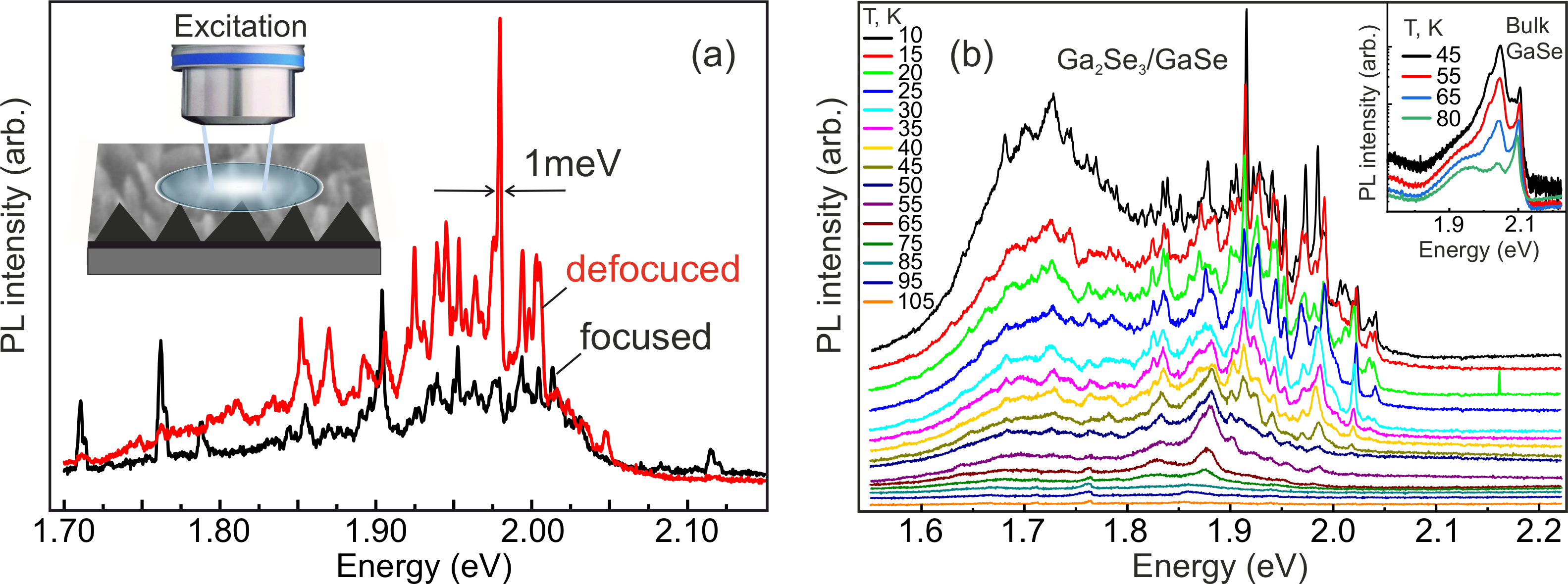}
\caption{(a) Narrow-line PL spectra recorded in a Ga$_2$Se$_3$/GaSe film with different excitation spot (focused to 1 $\mu$m and defocused). The excitation is illustrated by the inset against the bird-view SEM. Excitation power 0.2 mW, temperature 10 K. (b) Narrow-line PL spectra recorded at various temperatures in Ga$_2$Se$_3$/GaSe. For comparison, the inset shows the PL spectra in bulk GaSe, where the exciton peak begins to dominate with increasing temperature.
\label{figPLexc}}
\end{figure*}

\subsection{Transmission electron microscopy}

The coexistence of the allotropic $\epsilon$-GaSe and Ga$_2$Se$_3$ phases in the MBE-grown sample was confirmed by TEM studies of a thin flake (ridge) mechanically separated from the substrate (Fig.~\ref{figSR}c) shows both the $\epsilon$-GaSe hexagonal phase with a period of 0.8 nm and the Ga$_2$Se$_3$ phase with a period of $\sim$0.6 nm at the top of this flake. These phases, conjugated by their close-packed planes, have a small mismatch of about 2$\%$. The Ga$_2$Se$_3$ inclusions have, on average, a thickness of several monolayers and a transverse size of about 5 nm. In addition, hexagonal $\epsilon$-GaSe has a fairly high stacking-fault density, which is clearly seen from electron diffraction (not shown here).

According to previous theoretical studies \cite{Newman1962, Nakayama1997}, only two types of short-range ordering of vacancies are possible in Ga$_2$Se$_3$, which can occur either along [$\bar{1}$10] in ortho--Ga$_2$Se$_3$ or as a zigzag line along [$1\bar{1}$2] in mono--Ga$_2$Se$_3$ \cite{Nakayama1997}, although some uncertainty in the possible directions is still exists. In an orthorhombic cell $a$ = 5.477 \AA, $c$ = 5.414 \AA, while in a monoclinic structure $a$ = 6.66 \AA, $c$ = 11.65 \AA, i.e. the value of $c$ is twice as large \cite{Huang2013}. Therefore, in this flake we are most likely observing ortho-Ga$_2$Se$_3$. Note that such characterization using TEM is possible only at the thin tip of the trigonal flake, since the thicker part near the substrate is opaque for this method.

However, in this part, the stresses caused by the mismatch between the lattice parameters of the layer and the substrate are stronger, which can stimulate the formation of a monoclinic phase with a zigzag arrangement of vacancies, which is structurally more stable \cite{Huang2013}. Thus, the formation of both ortho- and mono-Ga$_2$Se$_3$ in the film under study cannot be ruled out.

\section{Narrow exciton lines and single-photon emission }

The possibility of quantum confinement in Ga$_2$Se$_3$ inclusions depends on their characteristic dimensions relative to the exciton Bohr radius R$_B$. In the first approximation, we can assume that it is close to R$_B$ = 4.5 nm in bulk GaSe \cite{Budweg}. In this case, the exciton diameter of 9 nm significantly exceeds the characteristic sizes of Ga$_2$Se$_3$ inclusions determined by the TEM method, and quantum confinement can provide the formation of distinct exciton levels, as in QDs. Therefore, the corresponding radiation must have a narrow radiation line width and a higher energy than the bulk material. The minimum energy of direct band gap transitions in bulk ortho-Ga$_2$Se$_3$ is about 1.75-1.85 eV \cite{Nakayama1997}. Although published data are strongly dispersed for mono-Ga$_2$Se$_3$, it is reliably established that its band gap is noticeably higher. 

The emission spectra measured in a grown film with a spatial resolution of $\sim$1 $\mu$m at low temperature (10 K) and low excitation power (0.2 mW) demonstrate ultra-narrow lines (Fig.~\ref{figPLexc}a). It should be emphasized that such narrow lines are registered in local places. With a slight shift, narrow lines quickly disappear (see Supplementary).  The ultra-narrow lines have a width of about 1 meV. Their intensity and number are increased as the excitation spot is defocused up to several $\mu$m. During laser defocusing, the direction of the incident light deviates slightly from the normal to the substrate, which is obviously preferable for the effective excitation of Ga$_2$Se$_3$ inclusions located inside inclined flakes (see the excitation diagram in Fig.~\ref {figPLexc}a). Under this condition, a weak peak at 2.12 eV corresponding to GaSe emission is also resolved. The region where ultranarrow lines appear (1.85-2.08 eV) indicates a possible combination of ortho (in the lower part) and mono (in the upper part) Ga$_2$Se$_3$ inclusions, for which the surrounding GaSe is a barrier.

At a higher excitation power (1 mW), a broad band appears centered around 1.73 eV (Fig.~\ref{figPLexc}b). This band is modulated by different peaks with a width of $\sim$3 meV. These peaks are probably due to the localization of excitons on structural defects, such as stacking faults, which were diagnosed by TEM in the film under study. Previously, we observed similar peaks in micro-PL spectra measured directly at stacking faults crossing ZnSe quantum wells \cite{Smirnov2018}. Such peaks and ultranarrow lines behave differently with increasing temperature: the ultranarrow lines disappear first in the upper region of the spectrum, since the GaSe barrier is not too high for Ga$_2$Se$_3$ inclusions, and broader peaks localized at deeper defect centers still exist. Interestingly, in bulk GaSe, the free exciton peak dominates in the PL spectra at elevated temperatures, although defect lines are present (see inset in Fig.~\ref{figPLexc}b).

An important argument in favor of our interpretation of ultranarrow lines can be their linear polarization. The large linear anisotropy of the transmission spectra in a vacancy-ordered bulk Ga$_2$Se$_3$ was first discovered in 1994 by Okamoto et al. \cite{Okamoto1994}. Later this phenomenon was theoretically considered by Nakayama and Ishikawa \cite{Nakayama1997} and P$\acute{e}$ressy et al. \cite{Peressi1998}. It is shown that optical transitions in Ga$_2$Se$_3$ must be strictly linearly polarized along certain crystal directions, where vacancies form a one-dimensional super-structure. The ordering of vacancies occurs along a line in the [$\bar{1}$10] direction in ortho--Ga$_2$Se$_3$ and in the form of a zigzag along [$1\bar{1}$2] in mono--Ga$_2$Se$_3$. With a more complex zigzag ordering, optical anisotropy may appear due to the off-diagonal component of the dielectric function tensor, which changes the main axis of the crystal towards the direction of vacancy ordering \cite{Nakayama1997}. As a result, a large optical anisotropy should be observed in measurements along and perpendicular to the corresponding direction in both mono- and ortho-Ga$_2$Se$_3$, without a shift in the PL line energy.

\begin{figure*}[t]
\includegraphics[width=\linewidth]{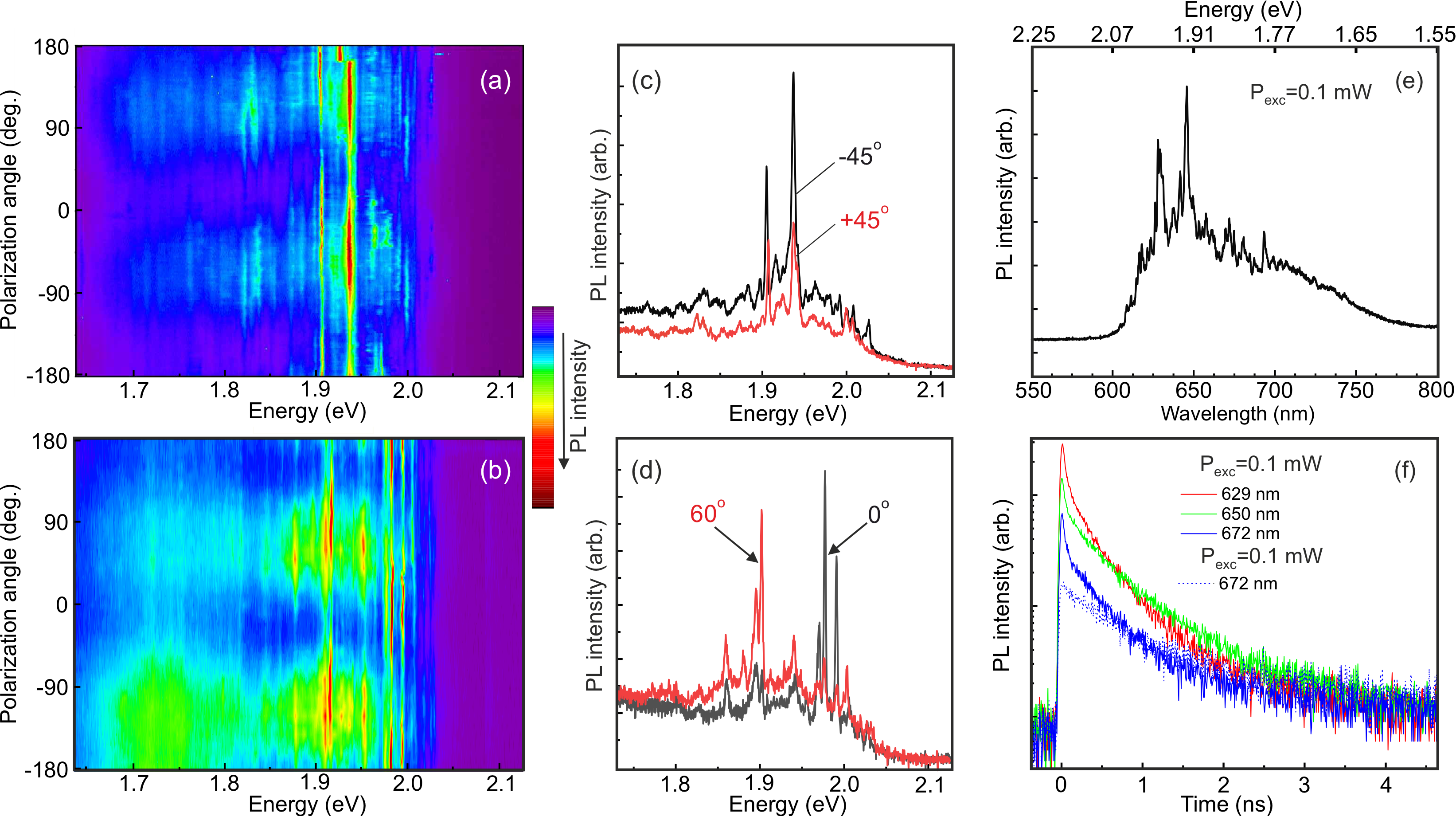}
\caption{(a, b) Mapping of the polarized spectra of narrow PL lines measured over the full planar angle from -180$^\circ$ to +180$^\circ$ in two characteristic points demonstrating the dominance of the ortho phase (a)  and the combination of the ortho and mono phases (b). The selected spectra demonstrate: (c) a difference of 90${^\circ}$ between the angles of maximum and minimum intensity in the ortho phase; (d) shows the spectra  characteristic of the ortho and mono phases. (e) Narrow-line PL spectrum and (f) PL decay curves of selected narrow lines measured at the indicated wavelengths at 10 K using a 405-nm pulsed-laser line. The excitation power is shown in the graphs.
\label{PolP12}}\end{figure*}

We have performed micro-PL polarization measurements within the full planar angle from -180$^{\circ}$ to +180$^{\circ}$, which confirm our findings.  A simultaneous change in the polarization of narrow lines in Figure ~\ref{PolP12}a indicates a strictly oriented crystallographic direction of the radiating objects located inside the excited spot. The difference in the angles of reaching the maximum and minimum of the PL intensity, $I$, is exactly 90$^{\circ}$, as shown in Fig.~\ref{PolP12}c. The degree of polarization, defined as 
$(I_{\bot}-I_{\parallel})/(I_{\bot}+I_{\parallel})$, varies from $\sim$40$\%$ to 60$\%$ for different narrow lines. We emphasize that such anisotropy is demonstrated by an ensemble of separate Ga$_2$Se$_3$/GaSe nanostructures, whereas earlier this phenomenon was studied in bulk Ga$_2$Se$_3$ \cite{Okamoto1993, Okamoto1994, Lovejoy2010}.

At some points, it turns out to be possible to register simultaneously two sets of lines, the polarization angle of which is shifted  with relative to each other by 60$^{\circ}$-90$^{\circ}$ (Fig.~\ref{PolP12} (b,d)). In one of the set, the lines of the strongest intensity are concentrated near 1.9 eV, and in the other, near 2.0 eV.  Accordingly, we attribute these sets to the radiation of the orthorhombic and monoclinic Ga$_2$Se$_3$ phases, which have the corresponding band gap energy for the bulk $\sim$1.75 eV and $\sim$1.9 eV. The higher-energy lines arise from mono-Ga$_2$Se$_3$ inclusions located closer to the substrate, where the stresses caused by the lattice misfit is stronger. The shift between the angles of polarization maxima in these two sets is due to the different crystal axes of the vacancy ordering in these two modifications of the defective zinc blende structure.  Thus, this is an important observation that confirms the existence of two structural phases of Ga$_2$Se$_3$/GaSe nanostructures implemented in the sample under study.

We studied the type of optical transitions (direct or indirect) corresponding to the narrow lines. Previously, by measuring the absorption and cw PL, it was shown that bulk Ga$_2$Se$_3$ has a direct band gap \cite{Park1989, Ho2014,Ho2020}. We assume that the measurements of time-resolved $\mu$-PL (TRPL) at low temperature is also one of the reliable methods, since in the case of a direct band structure the characteristic decay times should be much shorter than those of an indirect one. Figure~\ref{PolP12}(e-f) shows the measured decay curves of the ultra-narrow lines possessing different wavelengths. The 629-nm (1.97 eV) line probably comes from mono-Ga$_2$Se$_3$, and the following lines with wavelengths of 650 nm or more come from ortho-Ga$_2$Se$_3$. The characteristic decay times of the dominant component in all narrow lines are about 400 ps. Such short decay times are characteristic of direct-gap transitions. They cannot be attributed to GaSe, where indirect transitions dominate at low temperatures and have much longer decay times, on the order of tens of nanoseconds. For reference, we present in Supplementary the data on the PL decay in bulk GaSe. 
The presence of a weak and slowly decaying component in the decay curves is due to the admixture of radiation from the surrounding GaSe. We cannot prevent this, since the detection area of 1 $\mu$m significantly exceeds the dimensions of the Ga$_2$Se$_3$/GaSe nanostructures. However, at a lower excitation power (0.1 mW), when only the covering GaSe layer is effectively excited, the fast component from the inserts disappears. 



\begin{figure*}[t]
\centering
\includegraphics[width=1\linewidth]{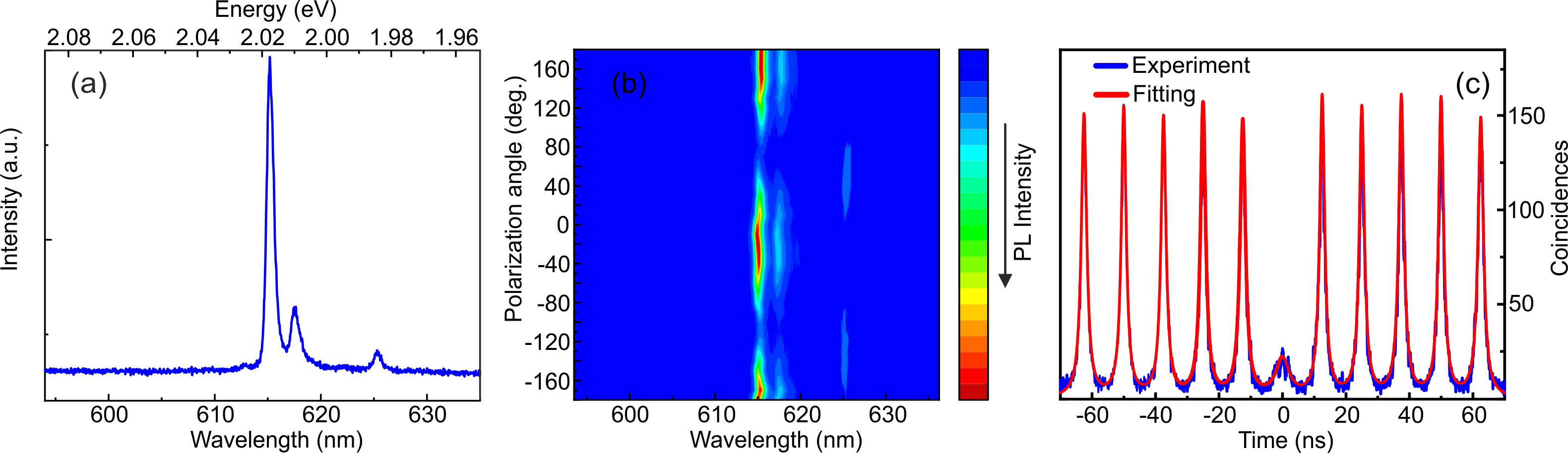}
\caption{(a) Spectrum of single excitonic line measured at 10 K. (b)  Mapping of the polarized spectra of single excitonic line measured over the full planar angle from -180$^\circ$ to +180$^\circ$. (c) Normalized second-order correlation function $g^{(2)}$ of single photon emission, measured at 10 K for a spectrally filtered single excitonic line. The obtained value of $g^{(2)}$(0) is 0.12.
\label{g2}}
\end{figure*}

Measurements of the photon statistics of the emission of narrow excitonic lines were carried out in a pulsed mode using a Hanbury Brown and Twiss setup. We focused on the part of the sample that exhibited the well-separated lines shown in Fig.~\ref {g2}a to highlight the brightest one. Polarization dependence mapping (Fig.~\ref {g2}b) shows that the selected line is presumably from mono-Ga$_2$Se$_3$, while weaker lines from the ortho phase are lower in energy and shifted by 60$^\circ$. The data array of measured photon statistics was processed as described in the Methods section. The resulting value at zero delay time corresponds to $g^{(2)}(0)\sim$0.12$\pm$0.01 (Fig.~\ref {g2}c). This value is well below the classical threshold of 0.5 used to prove single-photon emission. The observation of nonclassical light implies the existence of distinct quantum levels in Ga$_2$Se$_3$ inserts that act as QDs.

It is worth noting that the value $g^{(2)}(0)\sim$0.12 is much smaller than $g^{(2)}(0)=$0.37 reported in the paper \cite{Tonndorf2017} for defect-related single photon emitters in GaSe without coupling to a resonating waveguide. This coupling increases $g^{(2)}(0)$ to a value $\sim$0.13 \cite{TonndorfNano}, comparable to our results. Based on this trend, we can expect $g^{(2)}(0)$ to be close to zero for single-photon emitters using allotropic Ga$_2$Se$_3$/GaSe nanostructures when planarizing MBE-grown structures and forming a resonator on them.

\section{Conclusions}
Summarizing, this work presents Ga$_2$Se$_3$/GaSe nanostructures grown by van der Waals epitaxy. The specificity of these nanostructures is that they can hardly be called "hetero"-structures, since they are formed from related materials, consisting of the same atoms, but having different crystal structures. At a certain extent, they are allotropic compounds. These nanostructures exhibit ultranarrow exciton emission lines that persist in the PL spectra as long as the energy of their quantum levels does not exceed the height of the potential barrier, which is the surrounding GaSe matrix. The lines are strongly polarized with the degree of polarization up to 60$\%$. Taking into account the energy position of the narrow lines and their polarization, we assumed that the lines are associated with excitons confined in the two sets of Ga$_2$Se$_3$ inclusions which have the vacancy-ordered orthorhombic or monoclinic crystal structure. 
The time-resolved micro-PL spectroscopy of Ga$_2$Se$_3$/GaSe nanostructures established a direct type of narrow-line transitions with a corresponding alignment of the conduction and valence bands.  The non-classical nature of the narrow-line radiation was confirmed by the measurements of single-photon statistics. 
We should note that a similar ordering of vacancies has been observed in Ga$_2$S$_3$ and generally possible in other monochalcogenides such as In$_2$Se$_3$ and In$_2$Te$_3$. Therefore, we believe that our experiments will  pave the way for the use of allotropic nanostructures in quantum technologies.

\section{Methods}
\subsection{Van der Waals epitaxy}
A (Ga,Se)-based layer was grown on epi-ready GaAs(001) substrate using a two-chamber MBE setup (SemiTEq production). A standard Ga cell and Se valve cracking cell (Veeco, USA) with the cracking zone temperature $T_{Se}$(cr)=500$^\circ$C were used as molecular beam sources. To avoid problems with thermal cleaning of the substrate to remove the native oxide layers in the absence of an As flux, a GaAs buffer layer 200 nm thick was grown in a separate III-V chamber and then transferred to the III-VI chamber through vacuum. When the substrate temperature was stabilized at a targeted value of $T_{S}$=400$^\circ$C, the growth of GaSe was initiated with simultaneous opening the Ga and Se shutters.
After the growth of first $\sim$ 4 nm, the epitaxy temperature was increased up to the $T_{S}$$\sim$500$^\circ$C. The VI/III flux ratio was controlled by measuring the beam equivalent pressures (BEPs) of the corresponding elements at the substrate position by using a Bayard-Alpert ion gauge and maintained as high as $\sim$ 20 and $\sim$ 30 for the $T_{S}$=400$^0$C and 500$^\circ$C, respectively. Taking into account that the stoichiometric conditions on the growth surface correspond to $P_{Se}$/$P_{Ga}$ $\sim$ 12 and $\sim$ 25 at these growth temperatures \cite{Sorokin2020}, the actual VI/III flux ratio can be estimated as $\sim$ 1.2 and 1.6 for the first and second stages of growth, respectively.

\subsection{Structural characterization}
SEM characterization was performed using CamScan S4-90FE scanning electron microscope. The crystal structure of mechanically separated flakes was investigated by TEM studies using a Jeol JEM-2100F microscope. 

\subsection{Micro-Raman measurements}
Raman spectra were obtained in backscattering geometry at room temperature. For these investigations, we used a Horiba LabRAM HREvo UV-VIS-NIR-Open spectrometer (Horiba, Lille, France) with confocal optics. Olympus MPLN100x (Olympus, Tokyo, Japan) objective lens (NA = 0.9) focused the laser beam in a spot $\sim$1 $\mu$m in diameter. As an excitation source, a Nd:YAG laser (Torus Laser Quantum, Stockport, UK) with $\lambda$ = 532 nm  was used. To avoid damage and heating of the sample, the laser power on its surface was limited to 80 $\mu$W. A set of Bragg reflection and transmission filters was used to obtain Raman spectra in ultra-low-frequency range ($\omega$ < 50 cm$^{-1}$).

\subsection{Micro-photoluminescence measurements}
$\mu$-PL setup was used for optical properties investigation of GaSe structures. Non resonant optical excitation of a $cw$ laser (405 nm) was used for the $\mu$-PL measurements. Incident radiation was focused in 2-3 $\mu$m spot on the sample by an apoachromatic objective lens with NA of 0.42. The power density was ~ 4 W/cm$^2$. The collected emission was dispersed by a 0.5m monochromator with a 600/mm grating for detection $\mu$~PL spectra at selected wavelengths. For time-resolved $\mu$-PL studies the 405 nm line of a pulsed laser was used. Polarization control was achieved by means of a halfwave plate and a film polarizer installed in the detection channel.

\subsection{Photon correlation statistics}
Photon correlation measurements were performed in a Hanbury Brown-Twiss detection scheme exploiting two single-photon avalanche silicon diodes possessing the photon timing resolution of about 40 ps and a 50:50 beam splitter. The time intervals between the detection events are registered by an electronic start-and-stop scheme (SPC-130, Becker and Hickle). To determine the value of g$^{(2)}$(0), the area under all peaks of the experimental histogram was carefully calculated. To determine the area under the peaks, the conductive histogram was approximated by Lorentzian contours, after which each contour was integrated over all time delay values. The approximation took into account the correction of the background signal associated with dark readings of single-photon detectors, as well as with the background signal not associated with QD luminescence \cite{Lee2020}. The value g$^{(2)}$(0) results in the area under the peak at zero delay normalized by the area under the peak at non-zero delay averaged over conventional peak histograms.

\section*{Author Contributions}
The manuscript was written using contributions of all authors. In particular, M.V.R., A.I.G., I.A.E., V.Yu.D. and A.A.T. contributed to $\mu$-PL and Raman studies. S.V.S. contributed to MBE growth. D.A.K. contributed to TEM measurements. T.V.S. proposed a general physical model and supervised this work. All authors have given approval to the final version of the manuscript.

\section*{Conflicts of interest}
The authors declare no conflict of interest.

\section*{Acknowledgements}
The work of M.V.R., A.I.G. and I.A.E. was supported by a grant from the Russian Science Foundation (no. 22-22-20049, https://rscf.ru/project/22-22-20049/) and a grant from the St. Petersburg Science Foundation in accordance with agreement no. 21/2022 dated April 14, 2022.



\balance


\bibliography{rsc} 

\providecommand*{\mcitethebibliography}{\thebibliography}
\csname @ifundefined\endcsname{endmcitethebibliography}
{\let\endmcitethebibliography\endthebibliography}{}
\begin{mcitethebibliography}{55}
\providecommand*{\natexlab}[1]{#1}
\providecommand*{\mciteSetBstSublistMode}[1]{}
\providecommand*{\mciteSetBstMaxWidthForm}[2]{}
\providecommand*{\mciteBstWouldAddEndPuncttrue}
  {\def\EndOfBibitem{\unskip.}}
\providecommand*{\mciteBstWouldAddEndPunctfalse}
  {\let\EndOfBibitem\relax}
\providecommand*{\mciteSetBstMidEndSepPunct}[3]{}
\providecommand*{\mciteSetBstSublistLabelBeginEnd}[3]{}
\providecommand*{\EndOfBibitem}{}
\mciteSetBstSublistMode{f}
\mciteSetBstMaxWidthForm{subitem}
{(\emph{\alph{mcitesubitemcount}})}
\mciteSetBstSublistLabelBeginEnd{\mcitemaxwidthsubitemform\space}
{\relax}{\relax}

\bibitem[Bourdon \emph{et~al.}(1990)Bourdon, Bringuier, Portella, Vivi\'eres, and Piccioli]{Bourdon}
A.~Bourdon, E.~Bringuier, M.~Portella, M.~Vivi\'eres and N.~Piccioli, \emph{Physical Review Letters}, 1990, \textbf{65}, 1925\relax
\mciteBstWouldAddEndPuncttrue
\mciteSetBstMidEndSepPunct{\mcitedefaultmidpunct}
{\mcitedefaultendpunct}{\mcitedefaultseppunct}\relax
\EndOfBibitem
\bibitem[Guo \emph{et~al.}(2015)Guo, Xie, Li, Yang, Chen, Wang, Zhang, Andreev, Kokh, Lanskii, and Svetlichnyi]{Guo}
J.~Guo, J.-J. Xie, D.-J. Li, G.-L. Yang, F.~Chen, C.-R. Wang, L.-M. Zhang, Y.~M. Andreev, K.~Kokh, G.~Lanskii and V.~Svetlichnyi, \emph{Light: Science \& Applications}, 2015, \textbf{4}, e362\relax
\mciteBstWouldAddEndPuncttrue
\mciteSetBstMidEndSepPunct{\mcitedefaultmidpunct}
{\mcitedefaultendpunct}{\mcitedefaultseppunct}\relax
\EndOfBibitem
\bibitem[Sato \emph{et~al.}(2020)Sato, Tang, Watanabe, Ohsaki, Yamamoto, Tanabe, and Oyama]{Sato}
Y.~Sato, C.~Tang, K.~Watanabe, J.~Ohsaki, T.~Yamamoto, T.~Tanabe and Y.~Oyama, \emph{Journal of Physics Communications}, 2020, \textbf{4}, 1065007\relax
\mciteBstWouldAddEndPuncttrue
\mciteSetBstMidEndSepPunct{\mcitedefaultmidpunct}
{\mcitedefaultendpunct}{\mcitedefaultseppunct}\relax
\EndOfBibitem
\bibitem[Cai \emph{et~al.}(2019)Cai, Gu, Lin, Yu, Geohegan, and Xiao]{Cai}
H.~Cai, Y.~Gu, Y.-C. Lin, Y.~Yu, D.~Geohegan and K.~Xiao, \emph{Applied Physics Reviews}, 2019, \textbf{6}, 041312\relax
\mciteBstWouldAddEndPuncttrue
\mciteSetBstMidEndSepPunct{\mcitedefaultmidpunct}
{\mcitedefaultendpunct}{\mcitedefaultseppunct}\relax
\EndOfBibitem
\bibitem[Wang \emph{et~al.}(2020)Wang, Gao, Wei, Han, Wang, Gao, Liu, Han, and Zhang]{Wang2020}
Y.~Wang, J.~Gao, B.~Wei, Y.~Han, C.~Wang, Y.~Gao, H.~Liu, L.~Han and Y.~Zhang, \emph{Nanoscale}, 2020, \textbf{12}, 18356--18362\relax
\mciteBstWouldAddEndPuncttrue
\mciteSetBstMidEndSepPunct{\mcitedefaultmidpunct}
{\mcitedefaultendpunct}{\mcitedefaultseppunct}\relax
\EndOfBibitem
\bibitem[de~Vasconcellos \emph{et~al.}(2022)de~Vasconcellos, Wigger, Wurstbauer, Holleitner, Bratschitsch, and Kuhn]{Vasconcellos2022}
S.~M. de~Vasconcellos, D.~Wigger, U.~Wurstbauer, A.~W. Holleitner, R.~Bratschitsch and T.~Kuhn, \emph{Physica Status Solidi b}, 2022, \textbf{259}, 2100566\relax
\mciteBstWouldAddEndPuncttrue
\mciteSetBstMidEndSepPunct{\mcitedefaultmidpunct}
{\mcitedefaultendpunct}{\mcitedefaultseppunct}\relax
\EndOfBibitem
\bibitem[Tonndorf \emph{et~al.}(2017)Tonndorf, Pozo-Zamudio, Gruhler, Kern, Schmidt, Dmitriev, Bakhtinov, Tartakovskii, Pernice, de~Vasconcellos, and Bratschitsch]{TonndorfNano}
P.~Tonndorf, O.~D. Pozo-Zamudio, N.~Gruhler, J.~Kern, R.~Schmidt, A.~I. Dmitriev, A.~P. Bakhtinov, A.~I. Tartakovskii, W.~Pernice, S.~M. de~Vasconcellos and R.~Bratschitsch, \emph{Nano Letters}, 2017, \textbf{17}, 5446--5451\relax
\mciteBstWouldAddEndPuncttrue
\mciteSetBstMidEndSepPunct{\mcitedefaultmidpunct}
{\mcitedefaultendpunct}{\mcitedefaultseppunct}\relax
\EndOfBibitem
\bibitem[Belen'kii and Stopachinskii(1983)]{belenkii}
G.~L. Belen'kii and V.~B. Stopachinskii, \emph{Soviet Physics Uspekhi}, 1983, \textbf{26}, 497--517\relax
\mciteBstWouldAddEndPuncttrue
\mciteSetBstMidEndSepPunct{\mcitedefaultmidpunct}
{\mcitedefaultendpunct}{\mcitedefaultseppunct}\relax
\EndOfBibitem
\bibitem[Sarkar and Stratakis(2020)]{Sarkar2020}
A.~S. Sarkar and E.~Stratakis, \emph{Advanced Science}, 2020, \textbf{7}, 2001655\relax
\mciteBstWouldAddEndPuncttrue
\mciteSetBstMidEndSepPunct{\mcitedefaultmidpunct}
{\mcitedefaultendpunct}{\mcitedefaultseppunct}\relax
\EndOfBibitem
\bibitem[Sun \emph{et~al.}(2018)Sun, Luo, Zhao, Biswas, Li, and Zhang]{Sun2018}
Y.~Sun, S.~Luo, X.-G. Zhao, K.~Biswas, S.-L. Li and L.~Zhang, \emph{Nanoscale}, 2018, \textbf{10}, 7991--7998\relax
\mciteBstWouldAddEndPuncttrue
\mciteSetBstMidEndSepPunct{\mcitedefaultmidpunct}
{\mcitedefaultendpunct}{\mcitedefaultseppunct}\relax
\EndOfBibitem
\bibitem[Lai \emph{et~al.}(2022)Lai, Ju, Zhu, Wang, Wu, Yang, Zhang, Yang, Li, Cui, Deng, Han, Zhu, and Dai]{Lai2022}
K.~Lai, S.~Ju, H.~Zhu, H.~Wang, H.~Wu, B.~Yang, E.~Zhang, M.~Yang, F.~Li, S.~Cui, X.~Deng, Z.~Han, M.~Zhu and J.~Dai, \emph{Communications Physics}, 2022, \textbf{5}, 143\relax
\mciteBstWouldAddEndPuncttrue
\mciteSetBstMidEndSepPunct{\mcitedefaultmidpunct}
{\mcitedefaultendpunct}{\mcitedefaultseppunct}\relax
\EndOfBibitem
\bibitem[Capozzi and Montagna(1989)]{Capozzi1989}
V.~Capozzi and M.~Montagna, \emph{Physical Review B}, 1989, \textbf{40}, 3182\relax
\mciteBstWouldAddEndPuncttrue
\mciteSetBstMidEndSepPunct{\mcitedefaultmidpunct}
{\mcitedefaultendpunct}{\mcitedefaultseppunct}\relax
\EndOfBibitem
\bibitem[Ueno \emph{et~al.}(1997)Ueno, Takeda, Sasaki, and Koma]{Ueno1997}
K.~Ueno, N.~Takeda, K.~Sasaki and A.~Koma, \emph{Applied Surface Science}, 1997, \textbf{113}, 38--42\relax
\mciteBstWouldAddEndPuncttrue
\mciteSetBstMidEndSepPunct{\mcitedefaultmidpunct}
{\mcitedefaultendpunct}{\mcitedefaultseppunct}\relax
\EndOfBibitem
\bibitem[Liu \emph{et~al.}(2015)Liu, Yuan, Wang, Chen, Tang, Zhang, Zhang, Liu, Wang, Liu, Chen, Zou, Hu, and Xiu]{Liu2015}
S.~Liu, X.~Yuan, P.~Wang, Z.-G. Chen, L.~Tang, E.~Zhang, C.~Zhang, Y.~Liu, W.~Wang, C.~Liu, C.~Chen, J.~Zou, W.~Hu and F.~Xiu, \emph{ACS Nano}, 2015, \textbf{9}, 8592--8598\relax
\mciteBstWouldAddEndPuncttrue
\mciteSetBstMidEndSepPunct{\mcitedefaultmidpunct}
{\mcitedefaultendpunct}{\mcitedefaultseppunct}\relax
\EndOfBibitem
\bibitem[Pozo-Zamudio \emph{et~al.}(2015)Pozo-Zamudio, Schwarz, Sich, Akimov, Bayer, Schofield, Chekhovich, Robinson, Kay, Kolosov, Dmitriev, Lashkarev, Borisenko, Kolesnikov, and Tartakovskii]{DelPozo2015}
O.~D. Pozo-Zamudio, S.~Schwarz, M.~Sich, I.~A. Akimov, M.~Bayer, R.~Schofield, A.~Chekhovich, B.~J. Robinson, N.~Kay, O.~V. Kolosov, A.~I. Dmitriev, G.~Lashkarev, D.~N. Borisenko, N.~N. Kolesnikov and A.~I. Tartakovskii, \emph{2D Materials}, 2015, \textbf{2}, 035010\relax
\mciteBstWouldAddEndPuncttrue
\mciteSetBstMidEndSepPunct{\mcitedefaultmidpunct}
{\mcitedefaultendpunct}{\mcitedefaultseppunct}\relax
\EndOfBibitem
\bibitem[Li \emph{et~al.}(2023)Li, Zhang, Yang, Zhou, Song, Cheng, Zhang, Feng, Wang, Lu, Wu, and Chen]{Li2023}
W.~Li, X.~Zhang, J.~Yang, S.~Zhou, C.~Song, P.~Cheng, Y.-Q. Zhang, B.~Feng, Z.~Wang, Y.~Lu, K.~Wu and L.~Chen, \emph{Nature Communications}, 2023, \textbf{14}, 2757\relax
\mciteBstWouldAddEndPuncttrue
\mciteSetBstMidEndSepPunct{\mcitedefaultmidpunct}
{\mcitedefaultendpunct}{\mcitedefaultseppunct}\relax
\EndOfBibitem
\bibitem[Peressi and Baldereschi(1998)]{Peressi1998}
M.~Peressi and A.~Baldereschi, \emph{Journal of Applied Physics}, 1998, \textbf{83}, 3092--3095\relax
\mciteBstWouldAddEndPuncttrue
\mciteSetBstMidEndSepPunct{\mcitedefaultmidpunct}
{\mcitedefaultendpunct}{\mcitedefaultseppunct}\relax
\EndOfBibitem
\bibitem[Nakayama and Ishikawa(1997)]{Nakayama1997}
T.~Nakayama and M.~Ishikawa, \emph{Journal of the Physical Society of Japan}, 1997, \textbf{66}, 3887--3892\relax
\mciteBstWouldAddEndPuncttrue
\mciteSetBstMidEndSepPunct{\mcitedefaultmidpunct}
{\mcitedefaultendpunct}{\mcitedefaultseppunct}\relax
\EndOfBibitem
\bibitem[Okamoto \emph{et~al.}(1994)Okamoto, Yamada, Konagai, and Takahashi]{Okamoto1994}
T.~Okamoto, A.~Yamada, M.~Konagai and K.~Takahashi, \emph{Journal of Crystal Growth}, 1994, \textbf{138}, 204--207\relax
\mciteBstWouldAddEndPuncttrue
\mciteSetBstMidEndSepPunct{\mcitedefaultmidpunct}
{\mcitedefaultendpunct}{\mcitedefaultseppunct}\relax
\EndOfBibitem
\bibitem[Ho and Chen(2014)]{Ho2014}
C.-H. Ho and H.-H. Chen, \emph{Scientific Reports}, 2014, \textbf{4}, 6143\relax
\mciteBstWouldAddEndPuncttrue
\mciteSetBstMidEndSepPunct{\mcitedefaultmidpunct}
{\mcitedefaultendpunct}{\mcitedefaultseppunct}\relax
\EndOfBibitem
\bibitem[Park \emph{et~al.}(1989)Park, Kim, Kim, Kim, Jeong, Lee, and Lee]{Park1989}
K.-H. Park, H.-G. Kim, W.-T. Kim, C.-D. Kim, H.-M. Jeong, K.-J. Lee and B.-H. Lee, \emph{Solid State Communications}, 1989, \textbf{70}, 971--974\relax
\mciteBstWouldAddEndPuncttrue
\mciteSetBstMidEndSepPunct{\mcitedefaultmidpunct}
{\mcitedefaultendpunct}{\mcitedefaultseppunct}\relax
\EndOfBibitem
\bibitem[Morley \emph{et~al.}(1996)Morley, von~der Emde, Zahn, Offermann, Ng, Maung, Wright, Fan, Poole, and Williams]{Morley1996}
S.~Morley, M.~von~der Emde, D.~R.~T. Zahn, V.~Offermann, T.~L. Ng, N.~Maung, A.~C. Wright, G.~H. Fan, I.~B. Poole and J.~O. Williams, \emph{Journal of Applied Physics}, 1996, \textbf{79}, 3196--3199\relax
\mciteBstWouldAddEndPuncttrue
\mciteSetBstMidEndSepPunct{\mcitedefaultmidpunct}
{\mcitedefaultendpunct}{\mcitedefaultseppunct}\relax
\EndOfBibitem
\bibitem[{von der Emde} \emph{et~al.}(1996){von der Emde}, Zahn, Ng, Maung, Fan, Poole, Williams, and Wright]{Emde1996}
M.~{von der Emde}, D.~Zahn, T.~Ng, N.~Maung, G.~Fan, I.~Poole, J.~Williams and A.~Wright, \emph{Applied Surface Science}, 1996, \textbf{104-105}, 575--579\relax
\mciteBstWouldAddEndPuncttrue
\mciteSetBstMidEndSepPunct{\mcitedefaultmidpunct}
{\mcitedefaultendpunct}{\mcitedefaultseppunct}\relax
\EndOfBibitem
\bibitem[Lovejoy \emph{et~al.}(2010)Lovejoy, Yitamben, Ohta, Fain, Ohuchi, and Olmstead]{Lovejoy2010}
T.~Lovejoy, E.~Yitamben, T.~Ohta, S.~Fain, F.~Ohuchi and M.~Olmstead, \emph{Physical Review B}, 2010, \textbf{81}, 245313\relax
\mciteBstWouldAddEndPuncttrue
\mciteSetBstMidEndSepPunct{\mcitedefaultmidpunct}
{\mcitedefaultendpunct}{\mcitedefaultseppunct}\relax
\EndOfBibitem
\bibitem[Huang \emph{et~al.}(2013)Huang, Abdul-Jabbar, and Wirth]{Huang2013}
G.-Y. Huang, N.~Abdul-Jabbar and B.~Wirth, \emph{Journal of Physics: Condensed Matter}, 2013, \textbf{25}, 225503\relax
\mciteBstWouldAddEndPuncttrue
\mciteSetBstMidEndSepPunct{\mcitedefaultmidpunct}
{\mcitedefaultendpunct}{\mcitedefaultseppunct}\relax
\EndOfBibitem
\bibitem[Toropov \emph{et~al.}(2020)Toropov, Evropeitsev, Nestoklon, Smirnov, Shubina, Kaibyshev, Budkin, Jmerik, Nechaev, Rouvimov, Ivanov, and Gil]{Toropov2020}
A.~Toropov, E.~Evropeitsev, M.~Nestoklon, D.~Smirnov, T.~Shubina, V.~Kaibyshev, G.~Budkin, V.~Jmerik, D.~Nechaev, S.~Rouvimov, S.~Ivanov and B.~Gil, \emph{Nano Letteters}, 2020, \textbf{20}, 158--165\relax
\mciteBstWouldAddEndPuncttrue
\mciteSetBstMidEndSepPunct{\mcitedefaultmidpunct}
{\mcitedefaultendpunct}{\mcitedefaultseppunct}\relax
\EndOfBibitem
\bibitem[Ohuchi \emph{et~al.}(1990)Ohuchi, Parkinson, Ueno, and Koma]{Ohuchi1990}
F.~Ohuchi, B.~Parkinson, K.~Ueno and A.~Koma, \emph{Journal of Applied Physics}, 1990, \textbf{68}, 2168--2175\relax
\mciteBstWouldAddEndPuncttrue
\mciteSetBstMidEndSepPunct{\mcitedefaultmidpunct}
{\mcitedefaultendpunct}{\mcitedefaultseppunct}\relax
\EndOfBibitem
\bibitem[Koma(1999)]{Koma1999}
A.~Koma, \emph{Journal of Crystal Growth}, 1999, \textbf{201}, 236--241\relax
\mciteBstWouldAddEndPuncttrue
\mciteSetBstMidEndSepPunct{\mcitedefaultmidpunct}
{\mcitedefaultendpunct}{\mcitedefaultseppunct}\relax
\EndOfBibitem
\bibitem[Yuan \emph{et~al.}(2015)Yuan, Tang, Wang, Chen, Zou, Su, Zhang, Liu, Wang, Liu, Chen, Zou, Zhou, Hu, and Xiu]{Yuan2015}
X.~Yuan, L.~Tang, P.~Wang, Z.~Chen, Y.~Zou, X.~Su, C.~Zhang, Y.~Liu, W.~Wang, C.~Liu, F.~Chen, J.~Zou, P.~Zhou, W.~Hu and F.~Xiu, \emph{Nano Research}, 2015, \textbf{8}, 3332\relax
\mciteBstWouldAddEndPuncttrue
\mciteSetBstMidEndSepPunct{\mcitedefaultmidpunct}
{\mcitedefaultendpunct}{\mcitedefaultseppunct}\relax
\EndOfBibitem
\bibitem[Sorokin \emph{et~al.}(2019)Sorokin, Avdienko, I.V.~Sedova, Yagovkina, Smirnov, Davydov, and Ivanov]{Sorokin2019}
S.~Sorokin, P.~Avdienko, D.~K. I.V.~Sedova, M.~Yagovkina, A.~Smirnov, V.~Davydov and S.~Ivanov, \emph{Semiconductors}, 2019, \textbf{53}, 1131--1137\relax
\mciteBstWouldAddEndPuncttrue
\mciteSetBstMidEndSepPunct{\mcitedefaultmidpunct}
{\mcitedefaultendpunct}{\mcitedefaultseppunct}\relax
\EndOfBibitem
\bibitem[Sorokin \emph{et~al.}(2020)Sorokin, Avdienko, Sedova, Kirilenko, Davydov, Komkov, D.D.~Firsov, and Ivanov]{Sorokin2020}
S.~V. Sorokin, P.~Avdienko, I.~Sedova, D.~Kirilenko, V.~Davydov, O.~Komkov, D.~D.D.~Firsov and S.~Ivanov, \emph{Materials}, 2020, \textbf{13}, 3447\relax
\mciteBstWouldAddEndPuncttrue
\mciteSetBstMidEndSepPunct{\mcitedefaultmidpunct}
{\mcitedefaultendpunct}{\mcitedefaultseppunct}\relax
\EndOfBibitem
\bibitem[Ohtake and Sakuma(2021)]{Ohtake2021}
A.~Ohtake and Y.~Sakuma, \emph{The Journal of Physical Chemistry C}, 2021, \textbf{125}, 11257--11261\relax
\mciteBstWouldAddEndPuncttrue
\mciteSetBstMidEndSepPunct{\mcitedefaultmidpunct}
{\mcitedefaultendpunct}{\mcitedefaultseppunct}\relax
\EndOfBibitem
\bibitem[Singh and Gupta(2022)]{Singh2022}
D.~K. Singh and G.~Gupta, \emph{Materials Advances}, 2022, \textbf{3}, 6142--6156\relax
\mciteBstWouldAddEndPuncttrue
\mciteSetBstMidEndSepPunct{\mcitedefaultmidpunct}
{\mcitedefaultendpunct}{\mcitedefaultseppunct}\relax
\EndOfBibitem
\bibitem[Komkov \emph{et~al.}(2020)Komkov, Khakhulin, Firsov, Avdienko, Sedova, and Sorokin]{Komkov2020}
O.~Komkov, S.~Khakhulin, D.~Firsov, P.~Avdienko, I.~Sedova and S.~Sorokin, \emph{Semiconductors}, 2020, \textbf{54}, 1198--1204\relax
\mciteBstWouldAddEndPuncttrue
\mciteSetBstMidEndSepPunct{\mcitedefaultmidpunct}
{\mcitedefaultendpunct}{\mcitedefaultseppunct}\relax
\EndOfBibitem
\bibitem[Dai and Ohuchi(1998)]{Dai1998}
Z.~Dai and F.~Ohuchi, \emph{Applied Physics Letters}, 1998, \textbf{73}, 966--968\relax
\mciteBstWouldAddEndPuncttrue
\mciteSetBstMidEndSepPunct{\mcitedefaultmidpunct}
{\mcitedefaultendpunct}{\mcitedefaultseppunct}\relax
\EndOfBibitem
\bibitem[Teraguchi \emph{et~al.}(1991)Teraguchi, Konagai, Kato, and Takahashi]{Teraguchi1991}
N.~Teraguchi, M.~Konagai, F.~Kato and K.~Takahashi, \emph{Journal of Crystal Growth}, 1991, \textbf{115}, 798--801\relax
\mciteBstWouldAddEndPuncttrue
\mciteSetBstMidEndSepPunct{\mcitedefaultmidpunct}
{\mcitedefaultendpunct}{\mcitedefaultseppunct}\relax
\EndOfBibitem
\bibitem[Teraguchi \emph{et~al.}(1991)Teraguchi, Kato, Konagai, Takahashi, Nakamura, and Otuska]{Teraguchi2}
N.~Teraguchi, F.~Kato, M.~Konagai, K.~Takahashi, Y.~Nakamura and N.~Otuska, \emph{Applied Physics Letters}, 1991, \textbf{59}, 567--569\relax
\mciteBstWouldAddEndPuncttrue
\mciteSetBstMidEndSepPunct{\mcitedefaultmidpunct}
{\mcitedefaultendpunct}{\mcitedefaultseppunct}\relax
\EndOfBibitem
\bibitem[Okamoto \emph{et~al.}(1995)Okamoto, Takegami, Yamada, and Konagai]{Okamoto1995}
T.~Okamoto, T.~Takegami, A.~Yamada and M.~Konagai, \emph{Japanese Journal of Applied Physics}, 1995, \textbf{34}, 5984--5988\relax
\mciteBstWouldAddEndPuncttrue
\mciteSetBstMidEndSepPunct{\mcitedefaultmidpunct}
{\mcitedefaultendpunct}{\mcitedefaultseppunct}\relax
\EndOfBibitem
\bibitem[Okamoto \emph{et~al.}(1993)Okamoto, Konagai, Kojima, Yamada, Takahashi, Nakamura, and Nittono]{Okamoto1993}
T.~Okamoto, M.~Konagai, N.~Kojima, A.~Yamada, K.~Takahashi, Y.~Nakamura and O.~Nittono, \emph{Journal of Electronic Materials}, 1993, \textbf{22}, 229--232\relax
\mciteBstWouldAddEndPuncttrue
\mciteSetBstMidEndSepPunct{\mcitedefaultmidpunct}
{\mcitedefaultendpunct}{\mcitedefaultseppunct}\relax
\EndOfBibitem
\bibitem[Nagel \emph{et~al.}(1979)Nagel, Baldereschi, and Maschke]{Nagel}
S.~Nagel, A.~Baldereschi and K.~Maschke, \emph{Journal of Physics C: Solid State Physics}, 1979, \textbf{12}, 1625\relax
\mciteBstWouldAddEndPuncttrue
\mciteSetBstMidEndSepPunct{\mcitedefaultmidpunct}
{\mcitedefaultendpunct}{\mcitedefaultseppunct}\relax
\EndOfBibitem
\bibitem[Schl\"uter(1973)]{Shluter}
M.~Schl\"uter, \emph{Il Nuovo Cimento}, 1973, \textbf{3}, 313\relax
\mciteBstWouldAddEndPuncttrue
\mciteSetBstMidEndSepPunct{\mcitedefaultmidpunct}
{\mcitedefaultendpunct}{\mcitedefaultseppunct}\relax
\EndOfBibitem
\bibitem[Segura(2018)]{Segura2018}
A.~Segura, \emph{Crystals}, 2018, \textbf{8}, 206\relax
\mciteBstWouldAddEndPuncttrue
\mciteSetBstMidEndSepPunct{\mcitedefaultmidpunct}
{\mcitedefaultendpunct}{\mcitedefaultseppunct}\relax
\EndOfBibitem
\bibitem[Hoff(1975)]{Hoff1975}
R.~M. Hoff, \emph{The Journal of Physical Chemistry C}, 1975, \textbf{53}, 17\relax
\mciteBstWouldAddEndPuncttrue
\mciteSetBstMidEndSepPunct{\mcitedefaultmidpunct}
{\mcitedefaultendpunct}{\mcitedefaultseppunct}\relax
\EndOfBibitem
\bibitem[Lei \emph{et~al.}(2013)Lei, Ge, Liu, Najmaei, Shi, You, Lou, Vajtai, and Ajayan]{Lei2013}
S.~Lei, L.~Ge, Z.~Liu, S.~Najmaei, G.~Shi, G.~You, J.~Lou, R.~Vajtai and P.~Ajayan, \emph{Nano Letters}, 2013, \textbf{13}, 2777--2781\relax
\mciteBstWouldAddEndPuncttrue
\mciteSetBstMidEndSepPunct{\mcitedefaultmidpunct}
{\mcitedefaultendpunct}{\mcitedefaultseppunct}\relax
\EndOfBibitem
\bibitem[Ho(2020)]{Ho2020}
C.-H. Ho, \emph{ACS Omega}, 2020, \textbf{5}, 18527--18534\relax
\mciteBstWouldAddEndPuncttrue
\mciteSetBstMidEndSepPunct{\mcitedefaultmidpunct}
{\mcitedefaultendpunct}{\mcitedefaultseppunct}\relax
\EndOfBibitem
\bibitem[Finkman \emph{et~al.}(1975)Finkman, Tauc, Kershaw, and Wold]{Finkman1975}
E.~Finkman, J.~Tauc, R.~Kershaw and A.~Wold, \emph{Physical Review B}, 1975, \textbf{11}, 3785\relax
\mciteBstWouldAddEndPuncttrue
\mciteSetBstMidEndSepPunct{\mcitedefaultmidpunct}
{\mcitedefaultendpunct}{\mcitedefaultseppunct}\relax
\EndOfBibitem
\bibitem[Kolodziejczyk \emph{et~al.}(92)Kolodziejczyk, Filz, Krost, Richter, and Zahn]{Kolodziejczyk1992}
M.~Kolodziejczyk, T.~Filz, A.~Krost, W.~Richter and D.~Zahn, \emph{Journal of Crystal Growth}, 92, \textbf{117}, 549--—553\relax
\mciteBstWouldAddEndPuncttrue
\mciteSetBstMidEndSepPunct{\mcitedefaultmidpunct}
{\mcitedefaultendpunct}{\mcitedefaultseppunct}\relax
\EndOfBibitem
\bibitem[Yamada \emph{et~al.}(1992)Yamada, Kojima, and Takanashi]{Yamada1992}
A.~Yamada, N.~Kojima and R.~Takanashi, \emph{Japanese Journal of Applied Physics}, 1992, \textbf{31}, L186\relax
\mciteBstWouldAddEndPuncttrue
\mciteSetBstMidEndSepPunct{\mcitedefaultmidpunct}
{\mcitedefaultendpunct}{\mcitedefaultseppunct}\relax
\EndOfBibitem
\bibitem[von~der Emde \emph{et~al.}(1995)von~der Emde, Zahn, Ng, Maung, Fan, Poole, Williams, and Wright]{Emde1995}
M.~von~der Emde, D.~Zahn, T.~Ng, N.~Maung, G.~Fan, I.~Poole, J.~Williams and A.~Wright, \emph{Applied Surface Science}, 1995, \textbf{104/105}, 575--579\relax
\mciteBstWouldAddEndPuncttrue
\mciteSetBstMidEndSepPunct{\mcitedefaultmidpunct}
{\mcitedefaultendpunct}{\mcitedefaultseppunct}\relax
\EndOfBibitem
\bibitem[Irwin \emph{et~al.}(1973)Irwin, Hoff, Clayman, and Bromley]{Irwin1973}
J.~Irwin, R.~Hoff, B.~Clayman and R.~Bromley, \emph{Solid State Communications}, 1973, \textbf{13}, 1531--1536\relax
\mciteBstWouldAddEndPuncttrue
\mciteSetBstMidEndSepPunct{\mcitedefaultmidpunct}
{\mcitedefaultendpunct}{\mcitedefaultseppunct}\relax
\EndOfBibitem
\bibitem[Newman(1962)]{Newman1962}
P.~Newman, \emph{Journal of Physics and Chemistry of Solids}, 1962, \textbf{23}, 19\relax
\mciteBstWouldAddEndPuncttrue
\mciteSetBstMidEndSepPunct{\mcitedefaultmidpunct}
{\mcitedefaultendpunct}{\mcitedefaultseppunct}\relax
\EndOfBibitem
\bibitem[Budweg \emph{et~al.}(2019)Budweg, Yadav, Grupp, Leitenstorfer, Trushin, Pauly, and Brida]{Budweg}
A.~Budweg, D.~Yadav, A.~Grupp, A.~Leitenstorfer, M.~Trushin, F.~Pauly and D.~Brida, \emph{Physical Review B}, 2019, \textbf{100}, 045404\relax
\mciteBstWouldAddEndPuncttrue
\mciteSetBstMidEndSepPunct{\mcitedefaultmidpunct}
{\mcitedefaultendpunct}{\mcitedefaultseppunct}\relax
\EndOfBibitem
\bibitem[Smirnov \emph{et~al.}(2018)Smirnov, Belyaev, Kirilenko, Nestoklon, Rakhlin, Toropov, Sedova, Sorokin, Ivanov, Gil, and Shubina]{Smirnov2018}
D.~Smirnov, K.~Belyaev, D.~Kirilenko, M.~O. Nestoklon, M.~V. Rakhlin, A.~Toropov, I.~Sedova, S.~Sorokin, S.~Ivanov, B.~Gil and T.~Shubina, \emph{Physica Status Solidi-Rapid Research Letters}, 2018, \textbf{12}, 1700410\relax
\mciteBstWouldAddEndPuncttrue
\mciteSetBstMidEndSepPunct{\mcitedefaultmidpunct}
{\mcitedefaultendpunct}{\mcitedefaultseppunct}\relax
\EndOfBibitem
\bibitem[Tonndorf \emph{et~al.}(2017)Tonndorf, Schwarz, Kern, Niehues, Pozo-Zamudio, Dmitriev, Bakhtinov, Borisenko, Kolesnikov, Tartakovskii, de~Vasconcellos, and Bratschitsch]{Tonndorf2017}
P.~Tonndorf, S.~Schwarz, J.~Kern, I.~Niehues, O.~D. Pozo-Zamudio, A.~I. Dmitriev, A.~P. Bakhtinov, D.~N. Borisenko, N.~N. Kolesnikov, A.~I. Tartakovskii, S.~M. de~Vasconcellos and R.~Bratschitsch, \emph{2D Materials}, 2017, \textbf{4}, 021010\relax
\mciteBstWouldAddEndPuncttrue
\mciteSetBstMidEndSepPunct{\mcitedefaultmidpunct}
{\mcitedefaultendpunct}{\mcitedefaultseppunct}\relax
\EndOfBibitem
\bibitem[Lee \emph{et~al.}(2020)Lee, Leong, Kalashnikov, Dai, Gandhi, and Krivitsky]{Lee2020}
J.~Lee, V.~Leong, D.~Kalashnikov, J.~Dai, A.~Gandhi and L.~A. Krivitsky, \emph{AVS Quantum Science}, 2020, \textbf{2}, 031701\relax
\mciteBstWouldAddEndPuncttrue
\mciteSetBstMidEndSepPunct{\mcitedefaultmidpunct}
{\mcitedefaultendpunct}{\mcitedefaultseppunct}\relax
\EndOfBibitem
\end{mcitethebibliography}
\bibliographystyle{rsc} 

\end{document}